\documentclass[twocolumn,showpacs,preprintnumbers,amsmath,amssymb,superscriptaddress]{revtex4}
\usepackage{graphicx}
\usepackage{dcolumn}
\usepackage{bm}
\usepackage{color}
\usepackage{subfigure}
\usepackage{amsbsy}

\begin{document}

\title{Impact of the Meissner effect on magnetic micro traps for
neutral atoms near superconducting thin films}
\author{D. Cano}
\author{B. Kasch}
\author{H. Hattermann}
\author{D. Koelle}
\author{R. Kleiner}
\author{C. Zimmermann}
\author{J. Fort\'{a}gh}

\affiliation{Physikalisches Institut, Eberhard-Karls-Universit\"at
T\"ubingen, CQ Center for Collective Quantum Phenomena and their
Applications, Auf der Morgenstelle 14, D-72076 T\"ubingen, Germany}

\begin{abstract}
We theoretically evaluate changes in the magnetic potential arising
from the magnetic field near superconducting thin films. An example
of an atom chip based on a three-wire configuration has been
simulated in the superconducting and the normal conducting state.
Inhomogeneous current densities within the superconducting wires
were calculated using an energy-minimization routine based on the
London theory. The Meissner effect causes changes to both trap
position and oscillation frequencies at short distances from the
superconducting surface. Superconducting wires produce much
shallower micro traps than normal conducting wires. The results
presented in this paper demonstrate the importance of taking the
Meissner effect into account when designing and carrying out
experiments on magnetically trapped neutral atoms near
superconducting surfaces.
\end{abstract}

\maketitle
\section{Introduction}

Atom chips with microfabricated and nanofabricated field-generating
elements are useful devices for the coherent manipulation of
ultracold atoms \cite{Fortagh:07}. A diversity of potentials with
high spatial resolution can be generated near the chip surface using
state-of-the-art fabrication technology of metals, semiconductors,
superconductors and optical waveguides. By means of these potentials
Bose-Einstein condensates and degenerate Fermi gases are routinely
prepared. Significant progress in coherent manipulation and
state-sensitive detection of single atoms has been achieved in
recent years \cite{Teper:06,Colombe:06}. Applications of atom chips
include atom interferometers \cite{Wang:05, Schumm:05, Jo:07,
Günther:07}, ultra sensitive magnetic field sensors
\cite{Fortagh:02, Wildermuth:05} and portable experimental systems
for quantum gases \cite{Du:04}.

A key role on atom chips is played by the atom-surface interactions,
such as undesirable spin-decoherence mechanisms \cite{Henkel:99} and
attractive Casimir-Polder forces \cite{Lin:04, Harber:05}. The
lifetime of magnetic traps near the chip surface is limited by
decoherence mechanisms produced by the near-field noise radiation
from thermally induced currents in conductive surfaces \cite{Lin:04,
Henkel:99, Jones:03, Rekdal:04}. Cooling the chip can reduce those
thermal currents and thus increase the lifetime of the magnetic
traps. An important increase in the lifetime of several orders of
magnitude is expected when the surface layer crosses the transition
from the normal to the superconducting state \cite{Hohenester:07}.
The achievement of such conditions promises a new class of
experiments with cold atoms integrated with nanostructured surfaces
that will allow coherent control over the atoms into the submicron
range. Exciting proposals for coupling atoms to nanodevices such as
mechanical oscillators \cite{Treutlein:07} or superconducting
circuits \cite{Singh:07} outline new perspectives on experimental
research at the interface between atomic and solid-state physics.

Although the usage of superconducting microstructures on atoms chips
was already proposed in 1995 by Weinstein and Libbrecht
\cite{Weinstein:95}, and later on in several theoretical studies
\cite{Scheel:05,Skagerstam:06,Hohenester:07}, the impact of the
Meissner effect \cite{London:50,Ketterson:99} on the magnetic
trapping potential has not been considered as yet. In this paper, we
show that the magnetic-field expulsion from the superconductor has
an important effect on the magnetic trap. Our theoretical results
will be essential for the proper design of superconducting atom
chips, a subject on which the first experimental results have
recently been reported \cite{Roux:08, Mukai:07}.

Accurate methods to calculate the potential landscape near the chip
surface is a prerequisite for cold-atom experiments on atom chips.
In the case of using superconducting wires, these calculations must
take into account the Meissner effect, which implies that the
intensity of induced or applied currents decays exponentially into
the interior of the superconductor with a penetration depth
$\lambda$, which is also the depth to which the magnetic field
penetrates the superconductor. Magnetic-field calculations are
especially important at short distances from the chip surface, where
atoms are not easily accessible by imaging methods and where the
Meissner effect can have an important impact on the trap parameters.

The main goal of this paper is to investigate how the Meissner
effect modifies magnetic traps near superconducting thin films. To
accomplish this task, magnetic fields and trap parameters such as
position, oscillation frequencies and trap depth were calculated in
simulated chips containing thin-film wires. Simulations were carried
out for the vortex-free superconducting state and for the normal
conducting state, and the differences between the two cases are
analyzed.

Current distributions in superconducting thin films were calculated
in the frame of the London theory \cite{London:50}. Despite its
fundamental nature, exact solutions of the London theory exist only
for trivial cases such as a single sphere or a single cylinder in a
homogeneous magnetic field \cite{London:50}. Numerical methods are
therefore necessary to calculate current density distributions in
thin-film microstructures. Brandt and Mikitik \cite{Brandt:00}
reported on how to obtain numerical solutions of the London theory
for strips with rectangular cross section in a perpendicular
homogeneous magnetic field and/or with applied electric current.
More general geometries can be solved using commercial programs
which, however, have severe limitations. For example, most of them
provide accurate solutions only if the thickness ${\it h}$ of the
thin film is similar to the penetration depth $\lambda$
\cite{Kapaev:96}. In the present paper we overcome this limitation
and provide an algorithm that provides accurate solutions of the
London theory by finding the current distribution that minimizes the
free energy. A similar minimization method \cite{Pardo:03} has been
used to obtain the magnetization curves of arrays of superconducting
strips in homogeneous magnetic fields. The numerical method
presented in this paper can solve more general geometries in
arbitrary inhomogeneous magnetic fields, including most of the
geometries that are typically present in atom chips.

\section{Magnetic confinement on an atom chip} \label{sec:equations}

\begin{figure}
\centerline{\scalebox{0.45}{\includegraphics{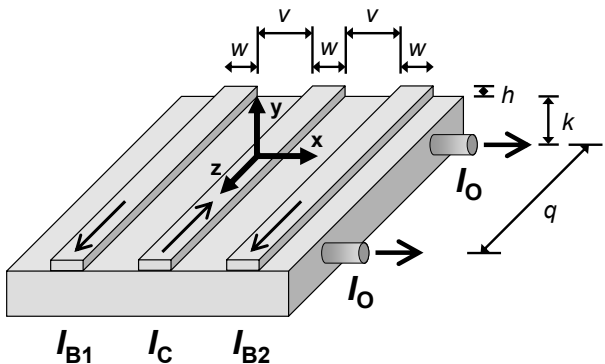}}}
\caption{Representation of a typical atom-chip geometry that
provides three-dimensional magnetic confinement \cite{Gunther:05}.
This theoretical example is used in this paper to study the
properties of magnetic micro traps near superconducting thin films.
Three parallel wires on the chip surface generate a two-dimensional
quadrupole field $\bm B_{2D}$ that provides radial confinement. The
width of the quadrupole wires and the separation between them are
denoted by $\it w$ and $\it v$, respectively. Underneath the chip
surface there are two offset wires in the perpendicular direction to
supply longitudinal confinement. The offset wires are located at
$z=q/2$ and $z=-q/2$.} \label{figchip}
\end{figure}

A magnetic micro trap can be realized with the atom-chip geometry
represented in Fig. \ref{figchip}. Three parallel thin-film wires on
the chip surface generate a two-dimensional confining field $\bm
B_{2D}$ (magnetic guide). The current $I_C$ in the central wire is
opposite in direction to the currents $I_{B1}$ and $I_{B2}$ in the
two outer wires. The magnetic guide forms at the position
$(x_0,y_0)$, where the magnetic field of the central current $I_C$
is canceled by the bias field generated by $I_{B1}$ and $I_{B2}$.
The field $\bm B_{2D}$ forms a two-dimensional quadrupole field
around $(x_0,y_0)$. Its modulus increases linearly in the radial
directions:
\begin{equation}
|\bm B_{2D}| = a \sqrt{(x - x_0)^2+(y - y_0)^2}. \label{B2D}
\end{equation}

Here, $a$ is the gradient of the magnetic guide in the radial
directions. For simplicity, these three parallel wires, which we
will refer to as quadrupole wires, are assumed to be identical in
size. The width and the thickness of each quadrupole wire are
denoted by $w$ and $h$, respectively. The outer wires are separated
from the central wire by $v$.

Longitudinal confinement is achieved by means of the inhomogeneous
offset field ${\bm B}_0$ generated by two offset wires perpendicular
to the quadrupole wires. The two offset wires are located below the
chip surface, and separated from the quadrupole wires by $k$. They
are driven with identical currents $I_0$.

The offset field ${\bm B}_0$ is superimposed onto the
two-dimensional confining field ${\bm B}_{2D}$, in such a way that a
magnetic trap forms between the two offset wires around the point
$(x_0, y_0,0)$. Near the centre of the trap,
\begin{equation}
{\bm B}_0 \simeq \left(
\begin{array}{c} 0 \\ a_0 z \\ b_0 + a_0
(y-y_0) + \frac{1}{2} b_z z^2
\end{array}\right),
\end{equation}
where $a_0$ and $b_z$ are the fist- and second-partial derivatives
of ${\bm B}_0$ with respect to the corresponding directions,
evaluated at $(x_0, y_0,0)$. The trap forms only if $b_z>0$; thus
for distances $y_0+k$ smaller than about 0.6 $q$. The offset field
$b_0$ at the trap center normally suffices to prevent Majorana
spin-flip transitions \cite{Sukumar:97} and the consequent loss of
atoms in Bose-Einstein condensates. However, in experiments with
thermal clouds at temperatures of several micro K, it might be
necessary to externally apply an additional homogeneous offset field
${\bm B}_{0,ext} = ( 0 , 0 , b_{0,ext})$.

The centre of the trap
\begin{equation}
\left( x_0 ,y_0-\frac{a_0(b_0+b_{0,ext})}{a_0^2+a^2},0 \right)
\end{equation}
is slightly displaced from the position $(x_0, y_0,0)$ of the
magnetic guide.

The offset field changes the radial potential from linear to
parabolic, with a harmonic oscillation parameter
$a^2/(b_0+b_{0,ext})$. The magnetic trap is then characterized by
the radial and longitudinal oscillation frequencies
\begin{equation}
\omega_r=\sqrt{\frac{g_F \mu_B m_F}{m (b_0+b_{0,ext})}} a
\hspace{2pt},\hspace{20pt} \omega_l=\sqrt{\frac{g_F \mu_B
m_F}{m}b_z} \hspace{2pt}. \label{formulafrequency}
\end{equation}
Here, $g_F$ is the Land\'{e} factor, $\mu_B$ is the Bohr magneton,
$m_F$ is the magnetic quantum number, and $m$ is the atom mass.

The $y$ component of ${\bm B}_0$ causes a small rotation of the
longitudinal axis of the magnetic trap. In the particular case that
the magnetic trap is in the plane $x$ = 0, and thus $I_{B1} =
I_{B2}$, the quadrupole field can be expressed as
\begin{equation}
{\bm B}_{2D} = \left(
\begin{array}{c} a (y-y_0) \\ a (x-x_0) \\ 0
\end{array} \right) \hspace{2pt},
\end{equation}

and the rotation occurs about the y axis \cite{Gunther:05}, with
angle $\theta=a_0/a$.

\section{Calculation of inhomogeneous current densities in
superconducting thin films} \label{sec:method}

Current densities are homogeneous in normally conducting wires.
Inhomogeneous current densities in the superconducting wires were
calculated numerically using the London theory. General solutions of
the London theory for the microstructure shown in Fig. \ref{figchip}
can be obtained by linear superposition of two separate cases. The
first of these describes the behavior of the microstructure when the
three quadrupole wires carry the respective currents $I_C$, $I_{B1}$
and $I_{B2}$. The second case describes the behavior of the
microstructure when each offset wire is driven with a current $I_0$
and no current is applied to the quadrupole wires. In the second
case, induced screening currents in the quadrupole wires can modify
the parameters of a magnetic trap formed near the chip surface.

For simplicity, the offset wires are assumed to be infinitely thin.
This approximation is valid if the width of the offset wires is much
smaller than $k$, in which case neither the screening currents in
the quadrupole wires nor the magnetic fields near the chip surface
depends on the current distribution in the offset wires. The
homogeneous offset field ${\bm B}_{0,ext} = ( 0 , 0 , b_{0,ext})$ is
not distorted by the Meissner effect because the longitudinal
demagnetizing factor of a strip quickly tends to zero as the strip
length increases to infinity \cite{Joseph:65}.

\subsection{Applied currents in the quadrupole wires}

\begin{figure}
\centerline{\scalebox{0.4}{\includegraphics{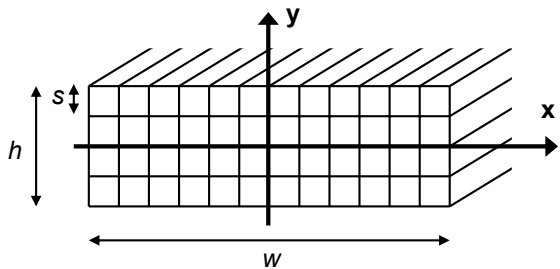}}} \caption{Sketch
of the division used to calculate the applied current distributions
in the quadrupole wires. A cross section of the central wire is
shown. Every wire is divided up into a large number of longitudinal
strips with squared cross section of side $s$. The current density
within each strip is assumed homogeneous.} \label{figstrips_scheme}
\end{figure}

Current density distributions in the superconducting quadrupole
wires are calculated with an energy-minimization numerical
procedure. Every quadrupole wire is divided up into a large number
$N$ of thin longitudinal strips with squared cross section of side
$s$ (see Fig. \ref{figstrips_scheme}). The current density is
assumed homogeneous within each thin strip, although it may vary
from strip to strip. The free energy of this system is the sum of
the magnetostatic energy of the currents and the kinetic energy of
the electrons \cite{Ketterson:99}, and can be written in the form
\begin{equation}
F(I_{\tt 1}, I_{\tt 2}, ... , I_{3N}) = \sum_{n=1}^{{\tt
3}N}\sum_{m=1}^{{\tt 3}N} I_n M_{nm} I_m, \label{energy}
\end{equation}

where $I_n$ is the electric current along the strip $n$ and $M_{nm}$
is the mutual inductance between the strips $n$ and $m$. The general
expression for $M_{nm}$ is \cite{Grover:46,Meservey:69}
\begin{eqnarray}
M_{nm} &=& \frac{\mu_0}{4 \pi} \int_{n}{\tt d^3}{\bm r}_n \int_{m}
{\tt d^3}{\bm r}_m \frac{J_n}{I_n} \frac{J_m}{I_m} \frac{1}{|{\bm
r}_n-{\bm r}_m|} + \nonumber \\ &+& \delta_{nm} \mu_0 \lambda^2
\int_{n} {\tt d^3}{\bm r}_n \frac{J_n^2}{I_n^2} \hspace{2pt},
\end{eqnarray}

where $\delta_{nm}$ is the Kronecker delta, $J_n=I_n/s^2$ is the
current density, and ${\bm r}_n$ denotes the position of point
within the strip $n$. The first and second terms represent the
magnetic and the kinetic inductances, respectively. The integrals
are carried out over the volumes of the corresponding strips. Since
the current density is homogeneous within each strip, the magnetic
term can be approximated by classical formulas tabulated in Ref.
\cite{Grover:46}. The integral of the kinetic term has a trivial
solution. The matrix elements then become
\begin{equation}
M_{nm} \simeq \left\{ \begin{array}{lr} \frac{\mu_0}{2 \pi} l \left(
\ln \frac{2 l}{d}-1 \right)& {\tt if} \hspace{5pt} n \neq  m
\hspace{2pt} , \\ \\ \frac{\mu_0}{2 \pi} l \left( \ln \frac{l}{s}+
\frac{1}{2} \right) + \mu_0 \frac{\lambda^2 l}{s^2} & {\tt if}
\hspace{5pt} n=m \hspace{2pt} , \end{array} \right.
\end{equation}

where $l$ is the length of the wires and $d$ is the distance between
the centers of the considered strips.

The superconducting current density is obtained by finding the set
of values $\{I_n\}_{n=1,\ldots,3N}$ that minimize the function
$F(I_{\tt 1}, I_{\tt 2}, ... , I_{3N})$. This is accomplished with
the method of the Lagrange multipliers, and imposing that the total
current flowing in each wire is fixed. The constraints are
\begin{equation}
\sum_{n=1}^{N} I_{n}=I_{\tt B1} \hspace{2pt} , \hspace{20pt}
\sum_{n=N+1}^{2N} I_{n}=I_{\tt C} \hspace{2pt} , \hspace{20pt}
\sum_{n=2N+1}^{3N} I_{n}=I_{\tt B2}.
\end{equation}

The equations to find the superconducting currents are
\begin{eqnarray}
\sum_{m=1}^{{\tt 3}N} M_{nm} I_m + \Lambda_1&=&0 \hspace{2pt} ,
\hspace{20pt} n=1,\ldots,N \hspace{2pt} , \nonumber
\\
\sum_{m=1}^{{\tt 3}N} M_{nm} I_m + \Lambda_2&=&0 \hspace{2pt} ,
\hspace{20pt} n=N+1,\ldots,2N \hspace{2pt} ,
\\
\sum_{m=1}^{{\tt 3}N} M_{nm} I_m + \Lambda_3&=&0 \hspace{2pt} ,
\hspace{20pt} n=2N+1,\ldots,3N \hspace{2pt} , \nonumber
\end{eqnarray}

where $\Lambda_1$, $\Lambda_2$ and $\Lambda_3$ are the Lagrange
multipliers. The solution of this system of linear equations is a
set of values $\{I_n\}_{n=1,\ldots,3N}$ that represent the current
distribution in the superconducting wires.

For long wires ($l>10w$) the calculated current distribution does
not depend on $l$. This limit is valid in all the examples shown in
this paper, where all wires are assumed infinitely long.

Low values of $s$ improve the accuracy of the solution at the
expense of long computation time. We have found that all values
lower than $\lambda/2$ produce practically the same numerical
results, and therefore, $s = \lambda / 2 $ is in general a good
choice for calculations.

Calculating the mutual inductance $M_{nm}$ between two strips is not
incompatible with the general definition of mutual inductance for
two closed circuits. One can imagine that every quadrupole wire is
part of a closed circuit that includes the current drivers and the
wires between the chip and the drivers. The mutual inductance
between two strips can be defined as the contribution of the two
strips to the total mutual inductance between the closed circuits of
which they form part.

\subsection{Screening currents in the quadrupole wires}

Screening currents arise in superconductors in the presence of
external magnetic fields as a consequence of the Meissner effect.
Screening currents can be calculated using the energy-minimization
method described in this section. This numerical method requires to
decompose the screening currents in small current elements
$\{I_n\}_{n=1,\ldots,N}$. In a magnetostatic situation, the
screening currents are closed, and therefore, a decomposition based
on small magnetic dipoles or small closed current elements is the
most appropriate.

\begin{figure}
\centerline{\scalebox{0.45}{\includegraphics{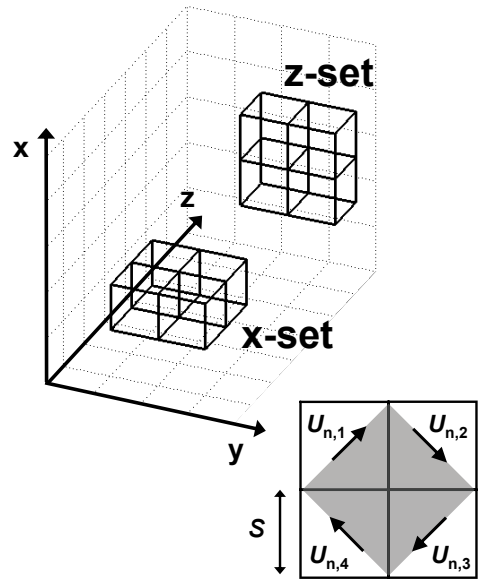}}}
\caption{Sketch of the sets of cubes in which the quadrupole wires
are divided to calculate the screening currents. The lower part
illustrates how the current $I_n$ is distributed within the set $n$.
The direction of the current, which is different in every cube, is
represented by four unit vectors $\bm{U}_{n,1}$, $\bm{U}_{n,2}$
$\bm{U}_{n,3}$ and $\bm{U}_{n,4}$. The effective surface of the set
is symbolized by a gray square of side $\sqrt{2} s$.}
\label{figcubos}
\end{figure}

The geometry of the closed current elements $I_n$ described in this
paragraph is suitable to evaluate the energy and the flux including
the kinetic term \cite{geometry}. The superconducting body is
divided up into small cubes of side $s$. The current is assumed to
be homogeneous within every small cube, although both the intensity
and the direction of the current might vary from cube to cube.
Closed current elements $I_n$ similar to magnetic dipoles are built
by grouping the cubes in sets of four, in the way illustrated in
Fig. \ref{figcubos}. The centers of the four cubes lie in the same
plane. Sets of four cubes can be built in planes parallel to the $x$
axis ($x$ sets), to the $y$ axis ($y$ sets) or to the $z$ axis ($z$
sets). In this manner, any cube that is not on the wire surface
belongs to twelve different sets: four $x$ sets, four $y$ sets and
four $z$ sets. Sets are sorted with numbers. Every set has
associated a current $I_n$, $n$ being the number of the set. The
current element $I_n$ is distributed within the set $n$ in a way
that is similar to a magnetic dipole, as shown in Fig.
\ref{figcubos}. The current $I_n$ changes its direction by 90
degrees as it passes from a cube to the next cube of the set. The
direction of $I_n$ in each cube is described by the unit vectors
${\bm U}_{n,1}$, ${\bm U}_{n,2}$, ${\bm U}_{n,3}$ and ${\bm
U}_{n,4}$.

The mutual inductance $M_{nm}$ between two sets $n$ and $m$ is
obtained by summing the contributions of the mutual inductances
between the separate cubes of both sets. Assuming one set $n$ made
up of the cubes $n_1$, $n_2$, $n_3$ and $n_4$ and another set $m$
made up of the cubes $m_1$, $m_2$, $m_3$ and $m_4$, the total mutual
inductance between the two sets is
\begin{equation}
M_{nm} = \sum_{a=1}^{\tt 4}\sum_{b=1}^{\tt 4} \hat{M}_{n_{a}m_{b}}
\left({\bm U}_{n,a} \cdot {\bm U}_{m,b}\right), \label{energy}
\end{equation}

where
\begin{eqnarray}
\hat{M}_{n_{a}m_{b}} &=& \frac{\mu_0}{4 \pi} \int_{n_a}{\tt d^3}{\bm
r}_{n,a} \int_{m_b}{\tt d^3}{\bm r}_{m,b} \frac{J_n}{I_n}
\frac{J_m}{I_m} \frac{1}{|{\bm r}_{n,a}-{\bm r}_{m,b}|} + \nonumber
\\ &+& \delta_{n_a m_b}\hspace{1pt}\mu_0\hspace{1pt}\lambda^2 \int_{n_a}
 {\tt d^3}{\bm r}_{n,a}
\frac{J_n^2}{I_n^2}
\end{eqnarray}

is the contribution of the cubes $n_a$ and $m_b$ to $M_{nm}$. The
scalar product ${\bm U}_{n,a} \cdot {\bm U}_{m,b}$ accounts for the
fact that this contribution depends on the angle between the current
directions. Here, ${\bm r}_{n,a}$ denotes the position of point
within the cube $n_a$, and $J_n=I_n/s^2$ is the current density in
the set $n$. The integrals are carried out over the volume of the
corresponding cubes. The double integral of the magnetic term was
calculated numerically \cite{Mathematica}, and the integral of the
kinetic term has a trivial solution. The matrix elements are then
approximated by
\begin{equation}
\hat{M}_{n_{a}m_{b}} \simeq \left\{ \begin{array}{lr}
\frac{\mu_0}{4 \pi} 1.88s+\frac{\mu_0 \lambda^2}{s} \hspace{2pt} , & d=0 \hspace{2pt} ,\\
\\
\frac{\mu_0}{4 \pi} 0.98s \hspace{2pt} , & d=s \hspace{2pt} ,\\
\\
\frac{\mu_0}{4 \pi} \frac{s^2}{d} \hspace{2pt} , & d>s \hspace{2pt}
,
\end{array} \right.
\end{equation}

where $d$ is the distance between cube centers.

For the reasons mentioned above, the mutual inductance between two
cubes is a senseless physical idea unless each cube is regarded as
part of a close circuit. To understand the meaning of
$\hat{M}_{nm}$, the screening current tubes can be regarded as a
collection of closed circuits with a certain inductance matrix. In
this way the mutual inductance between two cubes can be understood
as the contribution made by the cubes to the total mutual inductance
between the two current tubes in which the cubes are included. This
idea also applies to the self-inductance.

Every set of cubes is also characterized by the effective surface
${\bm S}_n$, which is represented in Fig. \ref{figcubos} by the gray
area. The effective surface is defined so that the product $M_{nm}
I_m$ is the flux produced by $I_m$ through ${\bm S}_n$. The modulus
of ${\bm S}_n$ is 2$s^2$, and its direction is determined by the
right-hand rule from the direction of the current $I_n$. Following
the notation of Fig. \ref{figcubos}, the effective surface can be
expressed by ${\bm S}_n = 2 s^2 {\bm U}_{n,1} \times{\bm U}_{n,2} $.

The solution of the London theory for this system is the current
distribution that cancels the flux -including both the magnetic and
the kinetic terms- in the interior of the superconducting wires. It
is possible to demonstrate \cite{London:50} that this condition is
equivalent to the free-energy minimization performed in the previous
section. The equations to be solved are formulated so that the flux
through every set of cubes is null:
\begin{equation} \sum_{m=1}^{N}
M_{nm} I_m + {\bm S}_n \cdot {\bm B}_n=0 \hspace{2pt} ,
\hspace{20pt} n=1,\ldots,N \hspace{2pt} \label{eqflux}.
\end{equation}

Here, $N$ is the total number of sets, $M_{nm}$ symbolizes the
elements of the inductance matrix, ${\bm B}_n$ is the external
magnetic field at the position of the set $n$, and ${\bm S}_n$ is
the effective surface of the set $n$. The first term in this
equation represents the total flux $\Phi_{S,n}$ induced by the
screening currents onto the set $n$. The second term is the magnetic
flux $\Phi_{0,n}$ of the external field onto the set $n$.

Due to its large size, the matrix $M_{nm}$ cannot be inverted using
any of the mathematic programmes that are generally available. As an
alternative, an iterative method has been used to solve Eq.
(\ref{eqflux}). In the fist step, a homogeneous screening current
distribution is assumed: $I_m^{(1)} = - {\bm S}_m \cdot {\bm B}_m /
M_{mm}$; $m=1,\ldots,N$. This distribution will not satisfy Eq.
(\ref{eqflux}), and the flux $\Phi_{S,n}^{(1)}=\sum_{m=1}^N M_{nm}
I_m^{(1)}$ created by the assumed screening currents onto each set
$n$ will not cancel the flux created by the offset wires
$\Phi_{0,n}$. In the second step, the current distribution is
calculated by $I_n^{(2)} = I_n^{(1)} - \xi_n^{(1)}(\Phi_{0,n}+
\Phi_{S,n}^{(1)})/M_{nn}$, which will generate a flux that will be
more similar to the desired solution. The process continues until
the convergence condition is satisfied:
$|\Phi_{0,n}+\Phi_{S,n}^{(e)}|< 10^{-4} \Phi_{0,n}$, after $e$
iterations. The number $\xi$ is a convergence factor that has no
physical meaning. This value is chosen by trial and error. The best
choice depends on the geometry of the superconducting body. If $\xi$
is too high, the method is not convergent; but if $\xi$ is too low,
the convergence is very slow. $\xi$ can vary from set to set and
also from step to step. The particular values of $\xi$ used for our
calculations will not be shown here since the results do not depend
on them and since there are many other possibilities.

Once the closed current elements $\{I_n\}_{n=1,\ldots,N}$ have been
obtained, the total current in every single cube is calculated by
summing the contributions of the sets of which the cube forms part.
From the total currents in the single cubes, the magnetic field
outside the superconducting body can be calculated.

The method presented in this section can be used to calculate
screening currents in superconducting thin-films induced by
arbitrary inhomogeneous magnetic fields. In the example shown in
Fig. \ref{figchip}, the magnetic field generated by the offset wires
has no component along the $x$ direction. For that reason, only $y$
sets and $z$ sets are used.

\section{Numerical results} \label{sec:results}

In this section we assess how the Meissner effect alters the
magnetic-trap parameters in the superconducting microstructure
depicted in Fig. \ref{figchip}. In all the examples shown in this
paper, the width of the quadrupole wires and the separation between
them are $w = 5 ~\mu$m and $v = 5 ~\mu$m, respectively. The
penetration depth $\lambda$ in the superconducting wires is 100 nm,
which is a typical value for metallic superconductors. The offset
wires are at a distance of $k = 5 ~\mu$m underneath the quadrupole
wires. The thickness of the quadrupole wires $h$ as well as the
separation between the two offset wires $q$ will be varied in the
examples in order to demonstrate their effect on the magnetic-trap
parameters.

\subsection{Quadrupole magnetic guide generated by the superconducting
chip}

\begin{figure}
\centerline{\scalebox{0.48}{\includegraphics{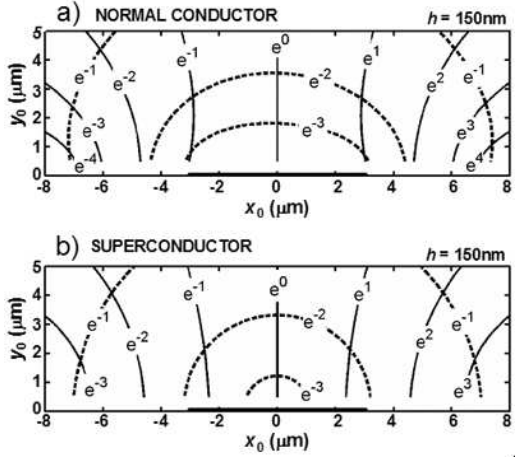}}}
\caption{Position $(x_0,y_0)$ of the magnetic guide in the
$x,y$-plane as a function of the ratios $\alpha$ and $\beta$ for the
superconducting and the normal conducting chip. Solid lines are
trajectories generated by varying $\alpha$ while keeping $\beta$
constant. Dashed lines are trajectories generated by varying $\beta$
while keeping $\alpha$ constant. The values of $\alpha$ and $\beta$
are written on the corresponding trajectory. The width and the
thickness of the quadrupole wires, and the separation between them
are $w = 5 \mu$m, $h = 150 \mu$m and $v = 5 \mu$m, respectively.}
\label{figbeta_alfa}
\end{figure}

First we analyze the position of the magnetic guide in the $x,y$
plane. The magnetic guide can be positioned within a large area
above the chip surface by changing the ratios
\begin{equation}
\alpha = \frac{I_C}{I_{B1} + I_{B2}} \hspace{3pt},\hspace{20pt}
\beta = \frac{I_{B1}}{I_{B2}} \hspace{2pt}.
\end{equation}
Figure \ref{figbeta_alfa} illustrates the trajectories corresponding
to constant $\alpha$ (dashed lines) and to constant $\beta$ (solid
lines) for the superconducting and the normal conducting chip.
Differences between the two cases are noticeable when the distance
between the magnetic trap and the chip surface is smaller than the
width of the wires.

\begin{figure}
\centerline{\scalebox{0.48}{\includegraphics{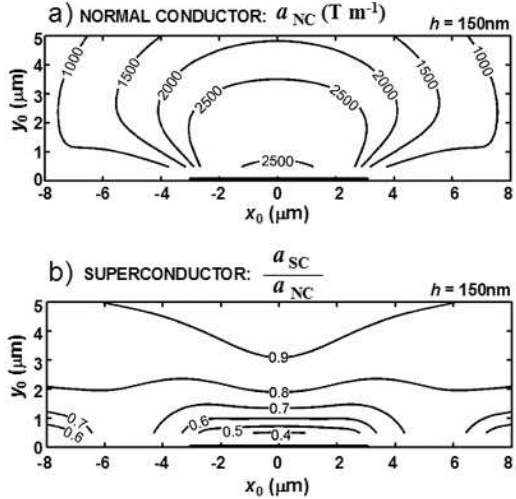}}}
\caption{Radial gradient obtained for different positions of the
magnetic guide $(x_0,y_0)$ keeping the sum of the currents $I_S =
I_C + I_{B1} + I_{B2}$ at a constant value of 1 A. a) Radial
gradient $a_{NC}$ in the normal conducting chip. b) ratios of the
gradient $a_{SC}$ in the superconducting chip to the gradient
$a_{NC}$ in the normal conducting chip. The width and the thickness
of the quadrupole wires, and the separation between them are $w = 5
\mu$m, $h = 150 \mu$m and $v = 5 \mu$m, respectively.}
\label{figgradients}
\end{figure}

In principle, both the position in the $x,y$ plane and the radial
gradient of the quadrupole field ${\bm B}_{2D}$ depend on the
applied currents $I_C$, $I_{B1}$ and $I_{B2}$. Once the ratios
$\alpha$ and $\beta$ have been chosen to position the magnetic
guide, the radial gradient can be varied by changing the value of
$I_S = I_C + I_{B1} + I_{B2}$. Fig. \ref{figgradients}(a) shows for
constant $I_S=$1 A the radial gradient $a_{NC}$ in the normal
conducting chip as a function of the position of the magnetic guide.
The radial gradient for other values of $I_S$ can be obtained by
linear scaling. For the superconducting case, the gradient $a_{SC}$
was calculated in the same way, keeping $I_S$ at a constant value of
1 A. Figure \ref{figgradients}(b) shows the ratio $a_{SC}/a_{NC}$.
Superconducting wires produce considerably lower radial gradients
than normal conducting wires. The radial gradient of ${\bm B}_{2D}$
is related with the radial oscillation frequency of the micro trap
by Eq. (\ref{formulafrequency}).

\begin{figure}
\centerline{\scalebox{0.4}{\includegraphics{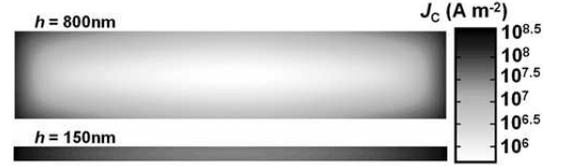}}}
\caption{Current density $J_c$ in the central wire for two different
values of the thickness $h$. The section of the wire is shown. The
width of the wires and the separation between them are $w = 5 \mu$m
and $v = 5 \mu$m, respectively. The applied currents are
$I_C$=65$\mu$A, and $I_{B1}$=$I_{B2}$=467$\mu$A ($\alpha$=0.07;
$\beta$=1).} \label{figcurrent_densities}
\end{figure}

Changes in the trapping field caused by the Meissner effect become
more pronounced when the superconducting wires are thicker or when
the penetration depth is smaller. Either thinner wires or longer
penetration depths imply more homogeneity in the superconducting
current densities, which results in magnetic fields which are more
similar to those produced by normal conductors. Figure
\ref{figcurrent_densities} shows the current-density distribution
$J_C(x,y)$ along the central wire for two different thicknesses.
Three regimes can be distinguished. If $h \gg \lambda$, the current
density decays exponentially from the surface and shows a sharp peak
in each corner. If $h \sim \lambda$, the current density becomes
homogeneous along the $y$ axis, having two maxima at $x = w/2$ and
$x = -w/2$. For extremely thin wires, the kinetic energy gets so
high that the current density becomes almost homogeneous, allowing
the magnetic flux to penetrate the film.

In the case of normal conducting wires, the magnetic-trap parameters
were independent of the thickness $h$. This is illustrated by
comparing the numerical results obtained for different values of
$h$. The variations in the $x,y$ position and in the radial gradient
produced by varying $h$ between 50$\mu$m and 800$\mu$m were,
respectively, less than 0.01$\mu$m and less than 0.1$\%$ at any
position within the area represented in Figs. \ref{figbeta_alfa} and
\ref{figgradients}. On the contrary, our numerical calculations
demonstrate that the magnetic-trap parameters depend considerably
upon the value of the thickness $h$ when the chip is
superconducting. For example, while for $h$=150 nm the radial
gradient in the superconducting chip is 0.4 times the radial
gradient in the normal conducting chip at 0.5 $\mu$m from the chip
surface above the central wire (see Fig. \ref{figgradients}), this
reduction factor is 0.6 for $h_1$=50 nm and 0.3 for $h_2$=500nm at
the same position. Therefore, the thickness of the thin-film wires
becomes relevant in the superconducting state.

\subsection{Longitudinal confinement in the superconducting state}

The analysis presented in this section is restricted to magnetic
traps located in the plane $x$ = 0. Different distances from the
chip surface will be considered.

\begin{figure}
\centerline{\scalebox{0.47}{\includegraphics{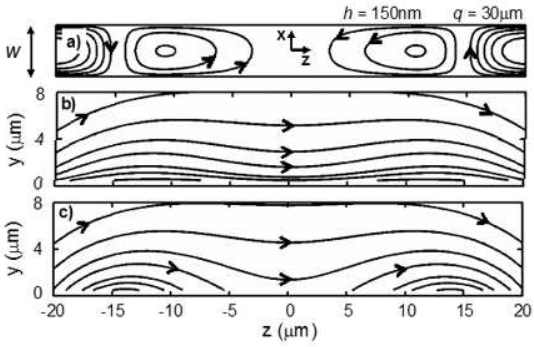}}}
\caption{Results obtained for the magnetic field ${\bm B}_0$ when
the offset wires are driven with equal currents $I_0$=1mA and no
current is applied to the quadrupole wires. a) stream lines of the
induced screening currents integrated along the $y$-direction in the
central wire. The plotted current density is 20$\mu$A/line. b) Field
lines in the plane $x=0$ above the superconducting chip. c) Field
lines in the plane $x=0$ above the normal conducting chip. The field
lines indicate the direction of the field but the density of lines
does not show the intensity of the field. One can appreciate the
expulsion of the magnetic field from the interior of the
superconducting wires. Calculations are performed for the following
geometrical parameters: $w=5\mu$m, $v=5\mu$m, $h$=150nm, $k=5 \mu$m
and $q = 30 \mu$m.} \label{figlineas}
\end{figure}

\begin{figure}
\centerline{\scalebox{0.5}{\includegraphics{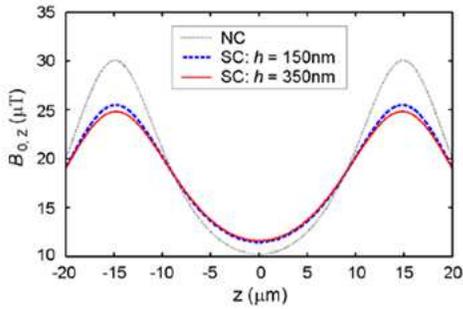}}}
\caption{Longitudinal component of the magnetic field along the
$z$-direction at $(x_0,y_0)=(0,2 \mu$m). Three different cases are
represented: normal conductor, superconductor with $h$ = 150nm, and
superconductor with $h$ = 350nm. $I_0$=1mA. The other geometrical
parameters are the same than in Fig. \ref{figlineas}.}
\label{figB0_z_150_350}
\end{figure}

Figure \ref{figlineas} shows the screening currents in the central
quadrupole wire as well as the magnetic field lines of the offset
field ${\bm B}_0$ for the superconducting and the normal conducting
chip. Figure \ref{figB0_z_150_350} represents the longitudinal
component of the magnetic field calculated along the $z$ direction
at $(x_0,y_0)=(0,2~\mu$m) for three different cases: normal
conductor, superconductor with $h$ = 150 nm, and superconductor with
$h$ = 350 nm. As with the results obtained for ${\bm B}_{2D}$,
differences between the superconducting and the normal conducting
states become larger with increasing $h$. The screening currents in
the superconducting quadrupole wires reduce the z component of the
magnetic field ${\bm B}_0$ at the positions $z=-q/2$ and $z=q/2$.
This effect entails a decrease of the trap depth along the
longitudinal direction. This reduction is of about 15\% at 2 $\mu m$
from the surface, and becomes higher than 25\% at distances of 1
$\mu m$ or shorter. For above than 10 $\mu m$, the reduction is
lower than 5\%.

\begin{figure}
\centerline{\scalebox{0.45}{\includegraphics{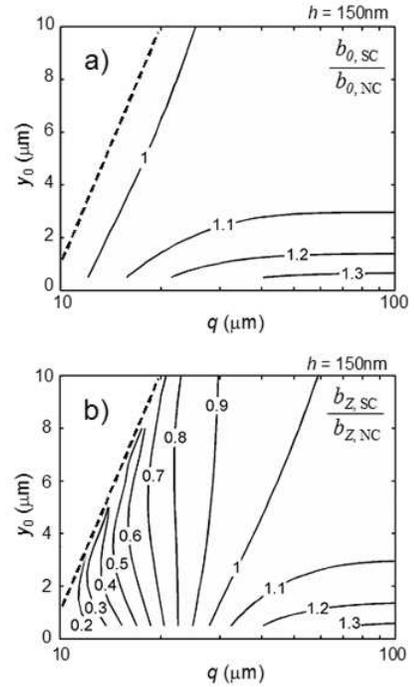}}} \caption{This
figure compares the trap parameters $b_0$ and $b_z$ between the
superconducting and the normal conducting chip. a) Ratio $b_{0,
SC}/b_{0, NC}$. b) Ratio $b_{z, SC}/b_{z, NC}$. The horizontal axis,
in logarithmic scale, represents the distance $q$ between the two
offset wires, and the vertical axis, in linear scale, represents the
position of the magnetic trap $y_0$. Data are represented in the
plane $x$=0. The other geometrical parameters are: $w=5\mu$m,
$v=5\mu$m, $h=150$nm, $k=5 \mu$m. The region in which no trap forms
is left of the dashed line.} \label{figb0_bz}
\end{figure}

The parameters $b_0$, $a_0$ and $b_z$ that describe the
inhomogeneous offset field ${\bm B}_0$ were numerically calculated
for the superconducting chip (SC) and the normal conducting chip
(NC) as a function of $q$ and $y_0$. Figure \ref{figb0_bz}(a) shows
the ratio $b_{0, SC}/b_{0, NC}$. The horizontal axis represents the
distance $q$ between the two offset wires, and the vertical axis
represents the position of the magnetic trap $y_0$. As observed in
this figure, the Meissner effect in the superconducting wires
slightly increases the value of $b_0$. This increase becomes more
significant as the distance between the two offset wires $q$ gets
longer, and the magnetic trap gets closer to the surface. Figure
\ref{figb0_bz}(b) shows the ratio $b_{z, SC}/b_{z, NC}$. The
longitudinal oscillation frequency of the microtrap is related to
$b_z$ by means of Eq. (\ref{formulafrequency}). As seen in Fig.
\ref{figb0_bz} the longitudinal frequency is dramatically reduced by
the Meissner effect when the offset wires are close to each other.
For high values of $q$ the effect is the opposite, and the
longitudinal frequencies are slightly higher in the superconducting
chip.

\begin{figure}
\centerline{\scalebox{0.45}{\includegraphics{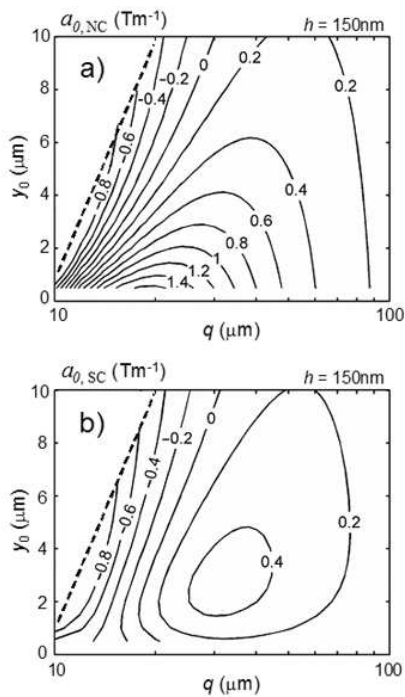}}} \caption{Trap
parameter $a_0$ in the normal conducting chip (a) and in the
superconducting chip (b). $I_0$=1mA. The horizontal axis, in
logarithmic scale, represents the distance $q$ between the two
offset wires, and the vertical axis, in linear scale, represents the
position of the magnetic trap $y_0$. Data are represented in the
plane $x$=0. The other geometrical parameters are: $w=5\mu$m,
$v=5\mu$m, $h=150$nm, $k=5 \mu$m. The region in which no trap forms
is left of the dashed line.} \label{figa0}
\end{figure}

Figure \ref{figa0} compares the value of $a_0$ between the
superconducting and the normal conducting chips. The parameter $a_0$
is related with the angle of rotation of the trap as explained in
Sec. \ref{sec:equations}. The calculated values of $a_0$ were
significantly lower in the superconducting microstructure than in
the normal conducting microstructure.

\section{Simulation of a magnetic micro trap}  \label{sec:simulation}

\begin{figure}
\centerline{\scalebox{0.4}{\includegraphics{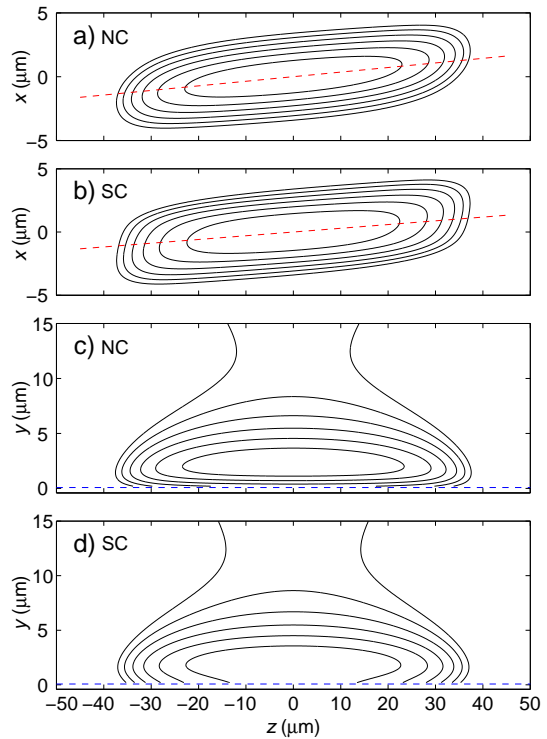}}}
\caption{Isopotential curves of a magnetic trap generated by the
atom chip shown in Fig. \ref{figchip} in the superconducting (SC)
and in the normal conducting (NC) state. The applied currents are
the same in both cases: $I_C = 0.2$mA, $I_{B1} = I_{B2} = 1.4$mA
($\alpha = 0.0714$, $\beta = 1$ and $I_S = 3$mA) and $I_0 = 2$mA.
Homogeneous offset field $b_{0,ext}=25\mu$T is externally applied to
stabilize the micro trap against Majorana losses. Following the
notation of Fig. \ref{figchip}, the geometry of the microstructure
is described by $w = 5 \mu$m, $v = 5 \mu$m, $h = 150$nm, $q = 100\mu
$m and $k = 5 \mu$m. The penetration depth is 100nm in the
superconducting wires. The two upper graphs show the isopotential
curves in the plane $y=y_0$. The dashed lines in the two upper
graphs represent the longitudinal axis of the micro trap, which is
rotated about the $y$-axis as explained before. The two lower graphs
show the isopotential curves in the plane perpendicular to the chip
surface along the longitudinal axis. The dashed lines in the two
lower graphs represent the chip surface. The parameters of this
micro trap are represented in Table 1. The magnetic field changes by
4$\mu$T per contour.} \label{figcontours_all}
\end{figure}

In this last section we apply the numerical results presented in
Sec. \ref{sec:results} to a typical example of a magnetic micro
trap. Figure \ref{figcontours_all} shows the isopotential curves of
a magnetic trap generated by the atom chip depicted in Fig.
\ref{figchip} in the superconducting and in the normal conducting
state. The applied currents are the same in both cases. The relevant
trap parameters are summarized in Table \ref{table}. The micro trap
forms closer to the surface in the superconducting chip than in the
normal conducting chip. In the present example, the Meissner effect
produces an important reduction in the radial oscillation
frequencies as well as a slight increase of the longitudinal
oscillation frequencies, as predicted by Figs. \ref{figgradients}
and \ref{figb0_bz}.

The most remarkable feature of the supercoducting chip is a
significant decrease of the trap depth towards the surface, which is
a consequence of the reduction of $a$ shown in Fig.
\ref{figgradients}. In the shown example, the trap depth is reduced
by about 80$\%$ in the superconducting chip.

\begin{table}[ht]
\begin{tabular}{l|c|c}
        & SC  & NC\\\hline
$a$ (Tm$^{-1}$)& 6.9 & 8.4 \\
$a_0$ (Tm$^{-1}$)& 0.2 & 0.3 \\
$b_0$ ($\mu$T)& 2.5 & 2.3 \\
$b_z$ (mTm$^{-2}$)& 5700 & 5240 \\
$y_0$ ($\mu$m)& 2.0 & 2.3 \\
$w_l$ (s$^{-1}$)& 2$\pi\cdot$ 95 & 2$\pi\cdot$ 92 \\
$w_r$ (s$^{-1}$)& 2$\pi\cdot$ 1650 & 2$\pi\cdot$ 2020 \\
$\theta$&1.7$^{\circ}$&2.1$^{\circ}$ \\
\end{tabular}
\caption{Parameters of the micro trap shown in Fig.
\ref{figcontours_all} for the superconducting (SC) and the normal
conducting (NC) states. Oscillation frequencies have been calculated
for $^{87}$Rb.}\label{table}
\end{table}

\section{Conclusion}

This theoretical study points out that differences between
superconducting and normal conducting chips become relevant when the
distance between the micro trap and the superconducting surface is
smaller than the width of the wires. The most dramatic consequence
of the Meissner effect is a significant reduction of the trap depth.
In general, the Meissner effect has to be taken into account when
designing and carrying out experiments with neutral atoms
magnetically trapped near superconducting surfaces. Although the
results shown in this paper have been obtained for the specific
example illustrated in Fig. \ref{figchip}, these conclusions can be
generalized to any atom chip made with superconducting thin films.

\begin{acknowledgments}
This work was supported by the DFG (SFB TRR 21) and by the BMBF
(NanoFutur 03X5506).
\end{acknowledgments}


\bigskip

\begin{thebibliography}{10}

\bibitem{Fortagh:07} J. Fort\'{a}gh, and C. Zimmermann, Rev. Mod. Phys. \textbf{79}, 235 (2007).
\bibitem{Teper:06} I. Teper, Y.-J. Lin, and V. Vuletic, Phys. Rev. Lett. \textbf{97}, 023002 (2006).
\bibitem{Colombe:06} Y. Colombe, T. Steinmetz, G. Dubois, F. Linke, D. Hunger, and J. Reichel, Nature  \textbf{450}, 272 (2007).
\bibitem{Wang:05} Y.~-J. Wang, D.~Z. Anderson, V.~M. Bright, E.~A. Cornell, Q. Diot, T. Kishimoto, M. Prentiss, R.~A. Saravanan, S.~R. Segal, and S. Wu, Phys. Rev. Lett. \textbf{94}, 090405 (2005).
\bibitem{Schumm:05} T. Schumm, S. Hofferberth, L.~M. Andersson, S. Wildermuth, S. Groth, I. Bar-Joseph, J. Schmiedmayer, and P. Kruger, Nat. Phys. \textbf{1}, 57 (2005).

\bibitem{Jo:07} G.-B. Jo, Y. Shin, S. Will, T. A. Pasquini, M. Saba, W. Ketterle, D.~E. Pritchard, M. Vengalattore, and M. Prentiss, Phys. Rev. Lett. \textbf{98}, 030407 (2007).
\bibitem{Günther:07} A. G\"{u}nther, S. Kraft, C. Zimmermann, and J. Fort\'{a}gh, Phys. Rev. Lett. \textbf{98}, 140403 (2007).
\bibitem{Fortagh:02} J. Fort\'{a}gh, H. Ott, S. Kraft, A. G\"{u}nther, and C. Zimmermann,  Phys. Rev. A \textbf{66}, 041604(R) (2002).
\bibitem{Wildermuth:05} S. Wildermuth, S. Hofferberth, I. Lesanovsky, E. Haller, L.~M. Andersson, S. Groth, I. Bar-Joseph, P. Kr\"{u}ger, and J. Schmiedmayer, Nature \textbf{435}, 440 (2005).
\bibitem{Du:04} S. Du, M.~B. Squires, Y. Imai, L. Czaia, R.~A. Saravanan, V.~M. Bright, J. Reichel, T.~W. H\"{a}nsch, and D.~Z. Anderson, Phys. Rev. A \textbf{70}, 053606 (2004).

\bibitem{Henkel:99} C. Henkel, S. P\"{o}tting, M. Wilkens, Applied Physics B \textbf{69}, 379 (1999).
\bibitem{Lin:04} Y.~J. Lin, I. Teper, C. Chin, and V. Vuletic, Phys. Rev. Lett. \textbf{92}, 050404 (2004).
\bibitem{Harber:05} D.~M. Harber, J.~M. Obrecht, J.~M. McGuirk, and E.~A. Cornell, Phys. Rev. A \textbf{72}, 033610 (2005).
\bibitem{Jones:03} M.~P.~A. Jones, C.~J. Vale, D. Sahagun, B.~V. Hall, and E.~A. Hinds, Phys. Rev. Lett. \textbf{91}, 080401 (2003).
\bibitem{Rekdal:04} P.~K. Rekdal, S. Scheel, P.~L. Knight, and E.~A. Hinds, Phys. Rev. A \textbf{70}, 013811 (2004).

\bibitem{Hohenester:07} U. Hohenester, A. Eiguren, S. Scheel, and E.~A. Hinds, Phys. Rev. A \textbf{76}, 033618 (2007).
\bibitem{Treutlein:07} P. Treutlein, D. Hunger, S. Camerer, T.~W. H\"{a}nsch, and J. Reichel, Phys. Rev. Lett. \textbf{99}, 140403 (2007).
\bibitem{Singh:07} M. Singh, arXiv:0709.0352v1 [quant-ph].
\bibitem{Weinstein:95} J.~D. Weinstein and K.~G. Libbrecht, Phys. Rev. A \textbf{52}, 4004 (1995).
\bibitem{Scheel:05} S. Scheel, P.~K. Rekdal, P.~L. Knight, and E.~A. Hinds, Phys. Rev. A \textbf{72}, 042901 (2005).

\bibitem{Skagerstam:06} Bo-Sture K. Skagerstam, U. Hohenester, A. Eiguren, and P.~K. Rekdal, Phys. Rev. Lett. \textbf{97}, 070401 (2006).
\bibitem{London:50} F. London, \textit{Superfluids} (Wiley, New York, 1950), Vol. I.
\bibitem{Ketterson:99} J.~B. Ketterson, and S.~N. Song, \textit{Superconductivity} (Cambridge University Press,1999), Chap. 2.
\bibitem{Roux:08} C. Roux, A. Emmert, A. Lupascu, T. Nirrengarten, G. Nogues, M. Brune, J.~-M. Raimond, and S. Haroche, arXiv:0801.3538v1 [physics.atom-ph].
\bibitem{Mukai:07} T. Mukai, C. Hufnagel, A. Kasper, T. Meno, A. Tsukada, K. Semba, and F. Shimizu, Phys. Rev. Lett. \textbf{98}, 260407 (2007).

\bibitem{Brandt:00} E.~H. Brandt, and G.~P. Mikitik, Phys. Rev. Lett. \textbf{85}, 4164 (2000).
\bibitem{Kapaev:96} M.~M. Khapaev, Supercond. Sci. Technol. \textbf{9}, 729 (1996).
\bibitem{Pardo:03} E. Pardo, A. Sanchez, and C. Navau, Phys. Rev. B \textbf{67}, 104517 (2003).
\bibitem{Gunther:05} A. G\"{u}nther, M. Kemmler, S. Kraft, C.~J. Vale, C. Zimmermann, and J. Fort\'{a}gh, Phys. Rev. A \textbf{71}, 063619 (2005).
\bibitem{Sukumar:97} C.~V. Sukumar, and D.~M. Brink, Phys. Rev. A \textbf{56}, 2451 (1997).

\bibitem{Joseph:65} R.~I. Joseph, and E. Schl\"{o}mann, J. Appl. Phys. \textbf{36}, 1579 (1965).
\bibitem{Grover:46} F.~W. Grover, \textit{Inductance Calculations} (D. Van Nostrand Company, Inc., New York,1946), Chap. 2\&5.
\bibitem{Meservey:69} R. Meservey, and P.~M. Tedrow, J.Appl.Phys. \textbf{40}, 2028 (1969).
\bibitem{Mathematica} It has been numerically solved in \textit{Mathematica} using the function \textit{NIntegrate}.
\bibitem{geometry} The energy and flux can be properly evaluated
from the closed current elements $\{I_n\}_{n=1,\ldots,N}$ when these
occupy the whole volume of the superconducting body and when the
overlapping between neighbor current elements is total, in the sense
that a homogeneous distribution of closed current elements generate
null electric current at any internal point that is not on the
surface of the superconducting body.


\end{thebibliography}
\end{document}